\documentclass[
superscriptaddress,
 amsmath,amssymb,
 aps,
pra,
twocolumn]{revtex4-2}
\usepackage{amsmath}
\usepackage{graphicx,stmaryrd}
\usepackage[utf8]{inputenc}
\usepackage[T1]{fontenc}
\usepackage{relsize}
\newcommand*\chem[1]{\ensuremath{\mathrm{#1}}}
\DeclareUnicodeCharacter{2212}{\ensuremath{-}}

\usepackage{dcolumn}
\IfFileExists{newtxtext.sty}
   {\usepackage{newtxtext,newtxmath}}
   {\IfFileExists{stix.sty}
      {\usepackage{stix}}
      {\IfFileExists{mathptmx.sty}
      {\usepackage{mathptmx}}{} } }
\usepackage{mathtools}
\usepackage{textcomp}
\usepackage{threeparttable, tablefootnote}

\usepackage{multirow}

\usepackage{hyperref}

\usepackage{xcolor}
\definecolor{LinkColor}{rgb}{0.256,0.439,0.588}

\hypersetup{
   colorlinks=true,
   linkbordercolor = {white},
   citecolor=LinkColor,
   linkcolor={black},
   urlcolor=LinkColor
   }
  
\begin{document}

\title{ Signature of T$_\textrm{c}$ above 111 K in Li-doped (Bi,Pb)-2223 superconductors: synergistic nature of hole concentration, coherence length and Josephson interlayer coupling }

\author{N.\ K.\ Man}
\affiliation{International Training Institute for Materials Science (ITIMS) and Faculty of Electronic Materials and Devices, School of Materials Science and Engineering, Hanoi University of Science and Technology, Hanoi 10000, Vietnam}

\author{Huu T.\ Do }
\email{huutdo09@gmail.com}
\affiliation{Department of Chemical Engineering, University of Illinois Chicago, Chicago, IL 60608, USA}

\date{\today}

\begin{abstract}

Understanding the bottleneck to drive higher critical transition temperature $T_\textrm{c}$ plays a pivotal role in the underlying study of superconductors.
We systematically investigate the effect of Li$^+$ substitution for Cu$^{2+}$ cations on the $T_\textrm{c}$, hole concentration, coherence length and interlayer coupling, and microstructure in Li-doped Bi$_{1.6}$Pb$_{0.4}$Sr$_2$Ca$_2$Cu$_3$O$_{10 + \delta}$ or (Bi,Pb)-2223 compound. 
Remarkably, we demonstrate by utilizing a long-time sintering accompanied by a multiple recurrent  intermediate stages of calcining and pressing within our renovated solid-state reaction method, the optimal Li-doped (Bi,Pb)-2223 samples achieve the well-enhanced $T_\textrm{c}$ of 111.1--113.8 K compared with the standard value of 110 K.
We evince the superconducting mechanism that the substitution of  Li$^{+}$ for Cu$^{2+}$ ions on the  CuO$_2$ layers causes augmenting the hole concentrations and promotes the correlation between the overdoped outer and the underdoped inner CuO$_2$ planes, and thus effects improve $T_\textrm{c}$.
Following a universal quadratic relation between $T_\textrm{c}$ and hole concentration, a new higher optimal hole concentration is provided. 
Additionally, by analyzing the Aslamazov-Larkin and Lawrence-Doniach theories on the reliable excess conductivity data near the critical temperature, we observe the strong effect of Li-doping on the system.
The coherence length steadily increases versus the Li-doped content, while the Josephson interlayer coupling strength between the CuO$_2$ layers almost remains a constant for the whole series of Li-doping.
 Our findings establish an insightful roadmap to improve the critical temperature and intrinsic superconducting properties in the Bi-2223 compounds through the doping process.

\end{abstract}



\maketitle

\section{Introduction}

In superconductors, the critical  temperature $T_\textrm{c}$ transforming from a non-superconducting to a superconducting state plays the key criterion for deciphering their fundamental properties and demonstrating potential for practical applications. 
Since the  high-$T_\text{c}$ superconductivity materials  containing copper oxide (referred to  cuprates)   was discovered 
 in the La-Ba-Cu-O compounds by  Bednorz and M\"{u}ller in 1986 ($T_\text{c} = 35$ K) \cite{Nobel-HTCS}, and right after that in the Y-Ba-Cu-O materials ($T_\textrm{c} = 93$ K) \cite{Keimer} which worked above the liquid nitrogen of 77 K, 
 there have been considerably theoretical, computational, and experimental efforts to elucidate how to obtain higher $T_\textrm{c}$.
Up-to-date, the cuprate family still holds the highest $T_\textrm{c}$ among all superconducting materials at ambient pressure.
The strategic idea to achieve the higher $T_\textrm{c}$ in the cuprate compounds  is to illuminate the superconducting  mechanism through intensive  investigation of electronic, antiferromagnetic and electron-phonon coupling properties on the CuO$_2$ layers, specifically the number of cuprate planes \cite{Science-2023-Bi2223,Rainer}, 
 their hole concentrations \cite{Keimer}, superconducting gap \citep{Ideta, PhysRevLett-2020-PRL}, and the interlayer coupling  between inequivalent cuprate layers \citep{Emery, Satoshi-PRL-2008, PhysRevB-O2-hole}.
 The $T_\textrm{c}$ trend of the cuprate compounds qualitatively pursues a well-established bell-shape as increasing number $n$ of CuO$_2$ layers in each unit cell and reaches  the maximum transition
temperature $T_\textrm{c}$ at $n = 3$ \cite{Keimer, Science-2023-Bi2223}.

\begin{figure}
\centering
\includegraphics[scale=0.465]{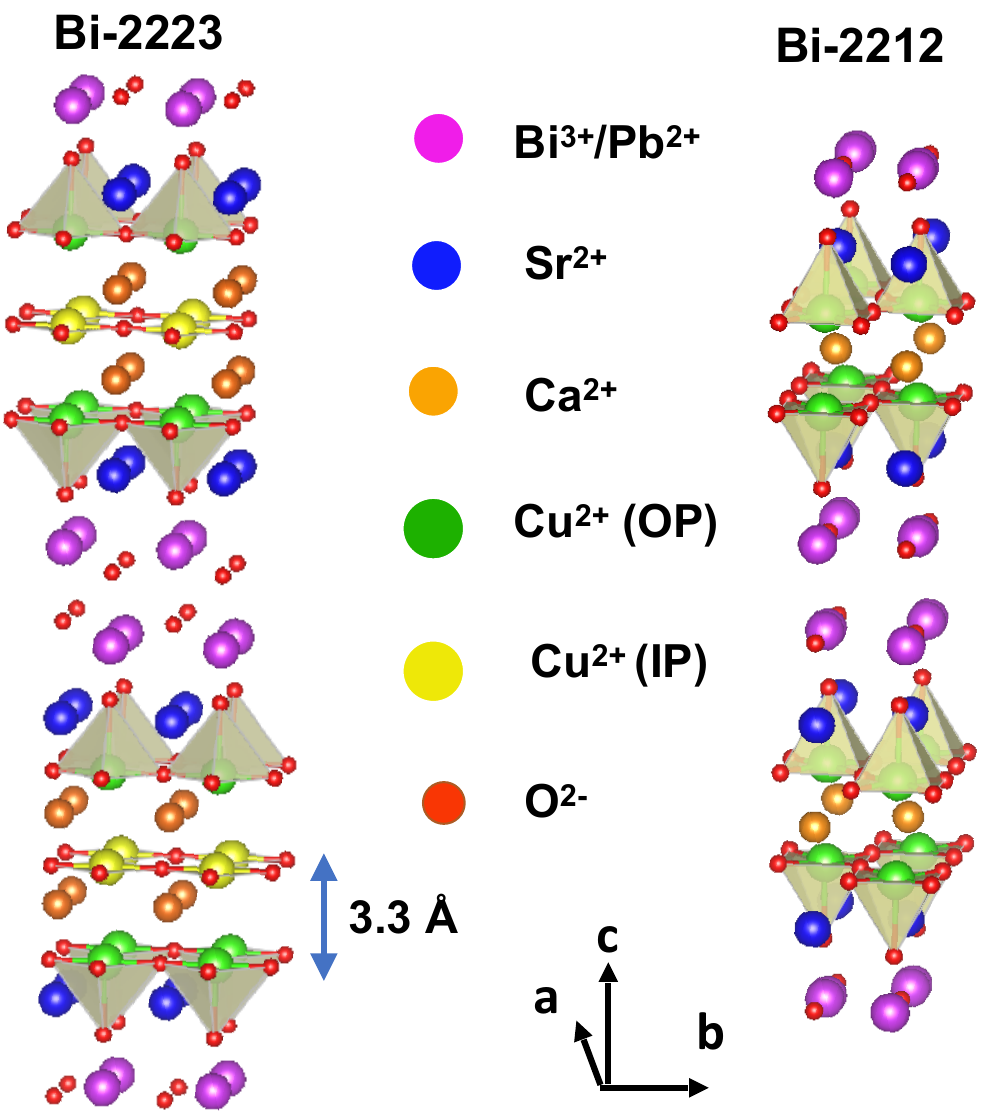}
\caption{ Illustrating the crystal structure of Bi-2223 and 2212 compounds. 
Bi-2212 consists of two homogeneous superconducting CuO$_2$ layers, which form a square pyramidal oxygen coordination, while Bi-2223 contains the two  square pyramids as outer layers sandwiching flat inner CuO$_2$ sheet.
(Bi,Pb)O, SrO and CaO are space layers. \label{fig:1}}
\end{figure}

Typically, a family of Bi$_2$Sr$_2$Ca$_{n-1}$Cu$_n$O$_{2n+4}$ (known as BSCCO) 
compounds show increasing $T_\text{c}$ with respect to the number $n$ of CuO$_2$ planes such as 36 K of Bi-2201 ($n$ = 1), 92 K of Bi-2212 ($n$ = 2), 110 K of Bi-2223 ($n$ = 3) and around 90 K for Bi-2234 ($n$ = 4) \citep{Keimer, Science-2023-Bi2223, Rainer}.
The BSCCO system consists of the superconducting CuO$_2$ layers and the space layers, including  Bi$_2$O$_3$, SrO, and CaO (see the illustrations of Bi-2212 and Bi-2223 in  Fig.~\ref{fig:1}).
 $T_\text{c}$  behaves analogously in the  dome trend as raising of the hole concentration on the CuO$_2$ layers to attain the maximum peak at an optimal hole level.
The optimal hole content for the Bi-2212 superconductors consisting of two homogeneous CuO$_2$ square pyramidal oxygen coordination (Fig.~\ref{fig:1}) is well-known to be 0.16 \cite{hole2-Bi2212, hole3-Bi2212}.
Through juxtaposing the trend of $T_\textrm{c}$  versus the hole concentration  in the double- and triple-layered single crystals,  
Piriou and {\it et al.}~implied a higher optimal hole value in the Bi-2223 single crystal than Bi-2212 due to higher oxygen pressure applied in the postannealing process   \cite{PhysRevB-O2-hole}.
Furthermore, it may deviate from the universal bell-shaped curve, in which  $T_\textrm{c}$  significantly raises in an underdoped (UD) regime, approaches the highest value at the optimal (OPT) point and slightly reduces in an overdoped (OD) region  \cite{PhysRevB-O2-hole, PhysRevLett-2020-PRL, NaturePhys-2023-Bi2223} or getting plateau in the OD  \cite{Fujii, B-Liang-2004}.
The controversial properties of Bi-2223 may be due to an anisotropic structure, distinct charge-transfer gap, and remarkable interlayer coupling between inner (IP) and outer (OP) planes  \cite{PhysRevB-O2-hole, PhysRevLett-2020-PRL, Science-2023-Bi2223}. 
 To our knowledge, 
 $T_\textrm{c}$ = 110.5--111 K typifies the highest value obtained in the Bi-2223 single crystal  by  magnetic susceptibility measurements \cite{PhysRevB-O2-hole, Clayton, B-Liang-2004}.

Since the hole concentration serves a primary  role in controlling the $T_\textrm{c}$ value in the Bi-2223 superconductors, several approaches are employed  to affect the parameter by adjusting the annealing oxygen environment \cite{PhysRevB-O2-hole, Clayton, Pb-doped-Bi2223, B-Liang-2004}, 
doping external elements to the system and performing different fabrication process such as conventional solid-state, sol-gel, thin-film, wire or single crystal \cite{PhysRevB-O2-hole, Clayton, CAO202412212, OH2019348, Apply-Phys-Let-2022, Super-Fluc-Bi2223, Bi-2223-wire,  nano13152197, Mg-doped-Bi2223}.
Doping impurity elements into the Bi-2223 system promises to alter the superconducting properties of the  CuO$_2$ layers as well as the critical temperature.
On the one hand, the direct substitution of transition metals such as Ni, Fe, and Co for Cu in the cuprate planes  reduces $T_\text{c}$ 
 and even suppresses  the hole concentration \citep{Pop, Pop1, Co-doped-Pop}.  
On the other hand,  monovalent alkaline  (Li, Na, K) or divalent elements (Mg, Ba) are common dopants in the  Bi-2223  superconductors and signify interesting properties \cite{Mg-doped-Bi2223}.
For instance,  Na$^+$-replacement for Ca$^{2+}$ ions in the insulating space layer increased $T_\textrm{c}$  while K-doping declined this parameter from 105.4 to 102.9 with 6\% at.\ for Sr$^{2+}$ ion \cite{OH2019348}.
In contrast,  substituting K$^+$ for Sr$^{2+}$ cations in the Bi-2223 compound increased $T_\textrm{c}$ from 107 to 110 K with 2\% at.~\cite{Wu-2007}. 
 Especially, Li$_2$O-added Bi-2223 materials showed a sign of an improvement for $T_\textrm{c} = 107$ to 108.1 K with a heavy doping of 20 \% at.\ in a whisker form,  but these values are not yet in the optimal regime of the Bi-2223 compounds with $T_\textrm{c} = 110.5$-111 K (see Fig.~\ref{fig:4}) \cite{Matsubara, MatsubaraI, PhysRevB-O2-hole, B-Liang-2004}.
Finding the optimal value of the hole concentration to enhance T$_\textrm{c}$ in the Bi-2223 superconductors through a doping process, particularly in the cuprate planes, still holds a challenging and thrilling question.

To our knowledge, the direct measurements  
 to access intrinsic quantities inside the cuprate layers such as a coherence length and interlayer coupling or even superexchange interaction of Cu-O-Cu to elucidate the relationship between the hole concentration and $T_\textrm{c}$  remains out of touch.
Performing the excess conductivity analysis  based on the Aslamazov and Larkin (AL) \cite{Aslamazov, ASLAMASOV1968238}, and Lawrence and
Doniach (LD) \cite{Law-Do} models in capable of providing for the quantitative estimations of the coherence length and Josephson interlayer coupling as well as studying superconducting fluctuation of the  Cooper pairs already creating at temperatures above the critical temperature \cite{NaturePhys-2023-Bi2223}. 
Moreover, the Josephson coupling between the IP and OP cuprate planes through an insulating CaO layer strongly impacts on determining the $T_\textrm{c}$, especially in trilayer Bi-2223 superconductors.  
The preceding thermodynamic fluctuation conductivity disclosed the effect of 
 cationic substitution on the critical temperature probably due to the interlayer coupling between
the CuO$_2$ layers or even enhancement of the coherence length in alkaline-doped Bi-2223 compound.
Despite intensively experimental studies have been implemented,
the origin of the Josephson interlayer coupling and the mechanism of $T_\textrm{c}$
advancement or destruction by external  replacements has not been
clearly unraveled  \cite{OH2019348, Fluct-Conduct2016, Super-Fluc-Bi2223, exc-conduct-Bi-2223}.

In this work, we investigate the superconducting properties and microstructures of the Li-doped 
(Bi,Pb)-2223, which was prepared using the upgraded solid-sate reaction method.  
We show that the $T_\textrm{c}$ values of our samples are greatly enhanced in the OPT regime by combining the long-time annealing process with multiple calcining and pressing steps.
Furthermore, as the Li-doped content increases, the hole concentration in the CuO$_2$ planes is advanced as well.
Consequently, the critical temperature reaches up to $113.8\pm 0.4$ K by the magnetic measurement, and this is the highest up-to-date  $T_\textrm{c}$ at ambient pressure in the Bi-2223 compounds. 
We organize our manuscript as follows. 
Sec.~\ref{Exp}  describes in detail our  experimental fabrication and measurement as well as the theoretical analysis of the excessive conductivity fluctuation. 
In Sec.~\ref{result}, we show and discuss the $T_\textrm{c}$ and  superconducting properties obtained by ac magnetic susceptibility and dc resistivity  artifacts.
 We enclosed our exploration in
Sec.~\ref{conclude}.

\section{Experimental details and model analysis \label{Exp}}

\subsection{Renovated solid-state reaction method and measuring approaches}

High-quality Li-doped (Bi,Pb)-2223 ceramic superconductors were performed with non-stoichiometric formula  \chem{Bi_{1.6}Pb_{0.4}Sr_2Ca_2(Cu_{1-x}Li_x)_3O_{10 + \delta}}, with $x = 0.0$, 0.025, 0.05,  0.10, and 0.15, and $\delta$ denotes the oxygen surplus content, by our renovated solid-state reaction technique.
High-purity precursors (99.9 \%) of \chem{Bi_2O_3}, \chem{PbO}, \chem{CuO}  and \chem{SrCO_3}, \chem{CaCO_3} and \chem{Li_2CO_3}  were weighted and mixed with the nominal compositions. 
Our synthesizing procedure delved into three principal phases. 
First, the desired precursor powders were well mixed and finely ground in an agate mortar and a pestle for 3 hours. 
Second, they were pressed into square-shaped pellets and calcined in the air at three intermediate steps with the temperatures of 650, 800, 820$^\text{o}$C (48 hours for each stage).
This process ensures the homogeneity of the samples. 
Compared with our traditional solid-state method, we only carried out a one-time intermediate stage of pressing and calcining at 800$^\text{o}$C for 24h \cite{ManP-Ag}, and we increased both the time and the number of pre-sintering preparations.
The achievement of these high-quality compounds is attributed to this factor.
Finally, these samples were annealed for 168 hours at the sintering temperature $T_\text{s} = 850^\text{o}$C to form the high-$T_\text{c}$  (Bi,Pb)-2223 crystalline phase with an additional low-$T_\text{c}$  (Bi,Pb)-2212 phase matrix.

Partial Pb-replacement for Bi 
is previously studied in synthesizing the (Bi,Pb)-2223 compound to enhance a phase
nucleation and growth formation mechanism through the secondary liquid phase of Ca$_2$PbO$_4$ and  (Ca,Sr)$_2$CuO$_3$  as well as to improve the homogeneity of the samples \cite{Doped-Pb, Doped-Pb1}. 
Our sintering temperature of 850 $^\text{o}$C was chosen slightly above the formation of Bi-2212--Bi-2223 transformation line at 840$^\text{o}$C \cite{Phase-diagram-Bi2223}
Although the lead-substitution for Bi reduced the sintering temperature, it diminishes the hole concentration in CuO$_2$ plane as well as $T_\textrm{c}$ recorded in several past works \cite{Hole-doped-PRB, Pb-doped-Bi2223, Pb-doped-Bi2223}.
 Nevertheless, adding Pb opens magnificent possibilities to fabricate the high-quality Bi-2223 compound  \cite{Pb-doped-Bi2223, B-Liang-2004}. 
The Li-doped effect further reduced the sintering temperature as the detailed investigation in Li-added Bi-2223 whiskers \cite{Matsubara}. 
So, we conducted the sintering process at $T_\text{s} = 850^\text{o}$C, lowering from 855$^\text{o}$C in our past study \cite{ManP-Ag},  for 168 hours accompanied with three additional intermediate steps promoted the  composition homogeneity  during the solid-state reaction and accomplish excellent (Bi,Pb)-2223 ceramic compounds (discuss in the next section).

We shaped specimens in square bars with the dimensions of $2\times 2 \times 12\text{mm}^3 $.
Subsequently, we employed the measurements by attaching Li-doped Bi-2223 specimens to the cold finger of a Helium closed-cycle system (CTI Cryogenic 8200) and lock-in amplifier. 
Our working system was cooled down and heated up in the range of 20--300 K. 
We conducted the temperature dependence of dc resistivity $\rho (T) $ and complex ac susceptibility $\chi (T) $ ( $\chi = \chi^{\prime} + i \chi^{\prime \prime}$ with in-phase $\chi^{\prime}$ and the out-of-phase $\chi^{\prime \prime}$ components) by advanced four-probes and lock-in amplifier techniques, respectively.
The  susceptibility measurement is powered by an external ac  with a magnetic field magnitude $H_\textrm{ac}$ and  a frequency $f$.
The surface morphology of the samples was investigated in detail using Jeol-5410-LV Scanning Electron Microscope (SEM).

\subsection{Thermodynamic conductivity fluctuation}

 The layered cuprate  superconductors are attributed to a short coherence length and high anisotropy, the thermodynamic fluctuations above the mean-field critical temperature $T_\textrm{c, mf}$, which is the peak of the first derivative of $d\rho/dT$.  
The thermodynamic fluctuation of excess conductivity  proposed by Aslamazov and Larkin \cite{Aslamazov, ASLAMASOV1968238} plays an essential role in probing the intrinsic superconducting features and determining the dimensionality of the materials 
 using a scaling approach: 
\begin{equation}
\centering
\Delta \sigma = A \epsilon^{-\lambda}. 
\label{eq0}
\end{equation}
Here, $\lambda$ denotes the critical component, and $A$ is a  constant obtained from experimental data. The reduced temperature is expressed by $\epsilon = (T - T_\textrm{c, mf})/T_\textrm{c, mf}$. 
 The $\lambda$ parameter associates with the conduction's dimensionality $d$ by a relation $\lambda = 2-d/2$,
and receives the specific values of 0.3, 0.5, 1, 3,  which represent the critical (CR), three-dimensional (3D), two-dimensional (2D), and short-wave (SW) fluctuations regimes, respectively.
The conductivity fluctuation is computed by:
\begin{equation}
\label{eq01}
    \Delta \sigma = \sigma (T) - \sigma_{\textrm{n}}(T) = \frac{1}{\rho(T)} - \frac{1}{\rho_{\textrm{n}}(T)}, 
\end{equation}
where $\rho(T)$ corresponds to the full-scale measuring resistivity and $\rho_{\textrm{n}}(T)$ is the normal resistivity behavior obtain above the pseudogap temperature $T^{\ast}$  \cite{ exc-conduct-Bi-2223}.  
The meaning of the constant $A$ relates to   
\begin{equation}
\label{eq1}
A =   
\begin{cases} 
\mathlarger{ \frac{e^2}{\hbar 32 \xi_c(0) } } & \textrm{for 3D fluctuation,}\\
\mathlarger{ \frac{e^2}{\hbar 16 d } } & \textrm{for 2D fluctuation.}
\end{cases}
\end{equation}
 $e$ denotes the charge of the electron, $\hbar$ is the reduced Planck’s constant, $\xi_c(0)$ stands for a zero-temperature coherence length along the c-axis, and $d$ displays the effective layer thickness of the 2D system.
Interestingly, the ratio $e^2/\hbar$ was defined as a unit of the quantized Hall conductance with the  well-known value \cite{Qunatum-conductant-PRL}: 
\begin{equation}
    \frac{e^2}{ \hbar} = \frac{2\pi}{25812.8099} (\textrm{S}).
\end{equation}

The AL theory was reformulated by Lawrence and Doniach (LD) in capable of accounting for layer superconducting polycrystals \cite{Law-Do}. 
The excess conductivity of the LD model is manifested by:
\begin{equation}
\label{eq2}
    \Delta \sigma = \frac{e^2}{16 \hbar d \epsilon} \Big ( 1+ \frac{4 \mathcal{J}}{\epsilon}   \Big )^{-1/2},
\end{equation}
 with the interlayer coupling expression of 
 \begin{equation}
 \label{J-coupling}
     \mathcal{J} = \frac{2 \xi_c(0)}{d}.
 \end{equation}
 Here, the $\xi_c(0)$ and $d$ are acquired from Eq.~(\ref{eq1}).
 In the context of the absent magnetic field in dc resistivity measurement, the cuprate materials primarily arise in coupling of the CuO$_2$ planes through the  
Josephson tunneling, especially in the Bi-2223 superconductors \cite{ PhysRevB-O2-hole, Conduct-Fluc}.

\section{Results and discussion \label{result}}

\subsection{Characterize the superconducting transition by ac magnetic susceptibility}

\begin{figure}
\centering
\includegraphics[scale=0.29]{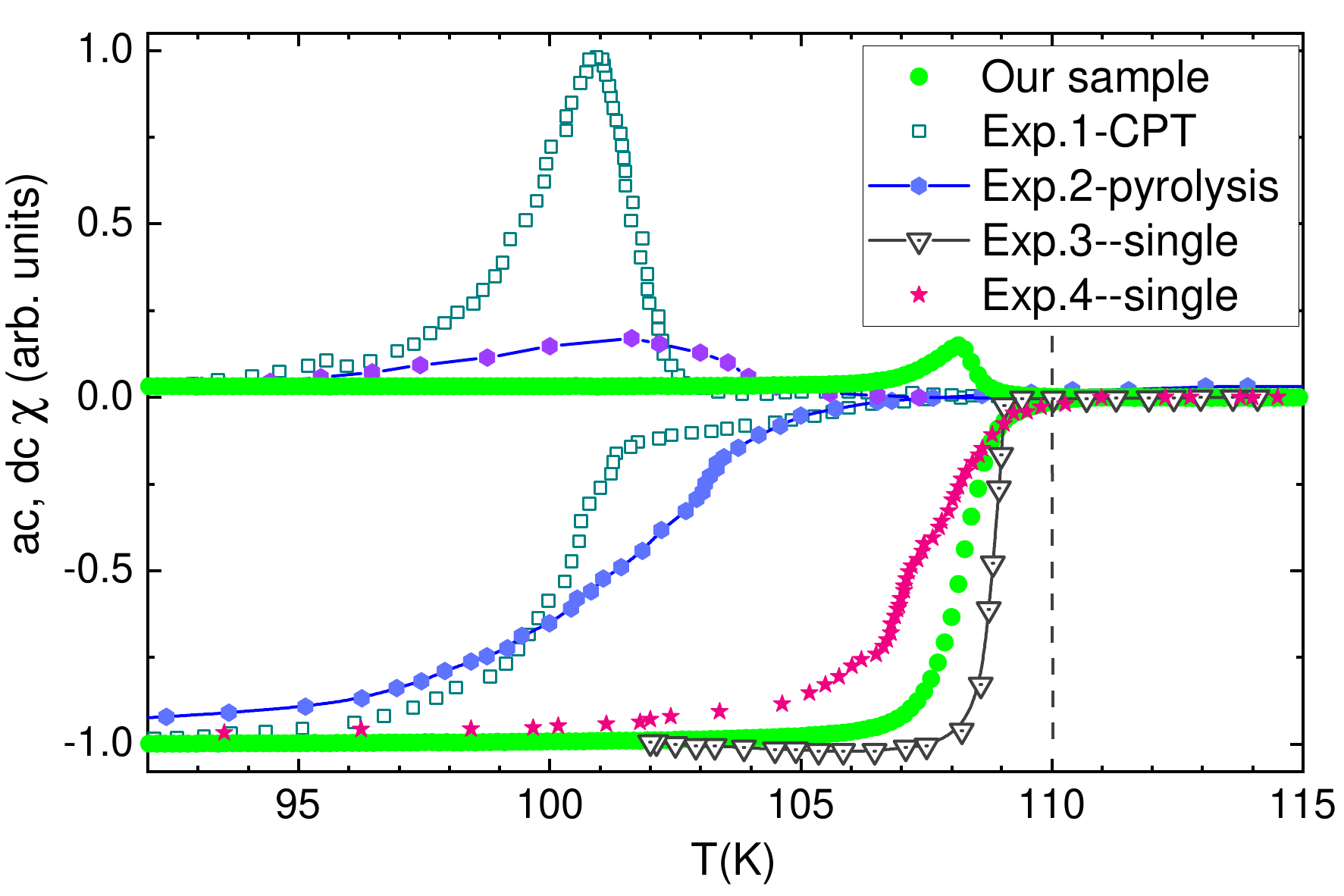}
\caption{ The magnetic susceptibility of our sample (green circle) with the chemical formula of  \chem{Bi_{1.6}Pb_{0.4}Sr_2Ca_2Cu_3 O_{10 + \delta}}  was synthesized by the upgraded solid-state reaction and measured at the field magnitude $H_\textrm{ac} = 2$ A/m  and frequency $f = 10$ kHz.  Exp.~1, 2, 3, and 4 data were extracted from Refs.~\cite{nano13152197, CAO202412212, Clayton, PhysRevLett-2020-PRL}, respectively. 
The detailed nominal composition, fabricating circumstances, and $T_\textrm{c,mag}$ are mentioned in Table~\ref{table0}. \label{fig:2}}
\end{figure}

\begin{table}
 \caption{ Summary of the $T_\textrm{c, mag}$ values and transition width $\Delta T$ for various Pb-doped and pure Bi-2223 superconductors near the optimal regime, prepared by different synthesizing techniques at the ac and dc (0 Hz in frequency) magnetic measurements. 
 The fabricating methods are noted with the concrete  conditions. 
 Here, ``solid'' refers to the solid-state reaction method;  
 ``single'' is the single-crystal growth, 
 ``CPT'' denotes for the chemical coprecipitation synthesis,
 and ``pyrolysis'' infers the spray-pyrolysis technique.  
 The nominal composition of all samples is  collected to Bi$_{2-y}$Pb$_{y}$Sr$_2$Ca$_2$Cu$_3$O$_{10 + \delta}$ formula in which $y$ denotes to lead-substituting content for Bi.  \label{table0}} 
 \begin{ruledtabular}
\begin{tabular}{l  l l l l l l l } 
 $y$ & $T_\textrm{c, mag}$  & $\Delta T$  & $H_\textrm{ac}$  & freq.  & method & annealing & Refs. \\ 
  & K & K & A/m & Hz &  & $^\circ$C, hours & \\
 \hline
 0.4  &  111.1 & 5 & 2 & 10$^4$  & solid & air, 850, 168  & ours  
\\
0 & 110.5 & 4 & 8 & 970  & single & O$_2$, 500, 240 &\cite{PhysRevB-O2-hole}
\\
0 & 109 & 3 & 27.9 & 0  & single &  O$_2$, 500, 100 & \cite{Clayton}
\\
0 & 111 & 12 & 800 & 0  & single & O$_2$, 500, 120 &\cite{B-Liang-2004}
\\
0 & 110 & 10 & N/A & 0  & single &O$_2$, N/A & \cite{PhysRevLett-2020-PRL}
\\
 0.4 & 107 & 5 & 39.9 & 219  & CPT & air, 850, 48 &\cite{nano13152197}
\\
 0.34 & 107 & 5 & 50 & 333  & solid & air, 860, 180 & \cite{Mg-doped-Bi2223}
\\
 0.34\footnote{Bi$_{1.8}$Pb$_{0.34}$Sr$_{2.1}$Ca$_{1.9}$Cu$_{3.07}$O$_{10 + \delta}$}  & 109.4 & 7.8 & 1600 & 0  & pyrolysis & O$_2$, 835, 78 & \cite{CAO202412212}
\\
\end{tabular}
\end{ruledtabular}
\end{table}

The magnetic response to an ac field is more susceptible to detect the phase transformation in superconductors than a resistivity and a dc static magnetization \cite{CELEBI1998131}.
When the superconducting material is cooled down at a low $H_\textrm{ac}$, 
the real part $\chi^{\prime}(T)$ deviates from a small positive susceptibility in the paramagnetic state to a bnn negative value within the diamagnetic state, and that signifies a magnetic critical temperature $T_\textrm{c, mag}$. 
At $T_\textrm{c, mag}$,  two electrons begin to collectively bind into a pair so-called the Cooper pair, which consequently triggers the expulsion of the external magnetic field \cite{Keimer}.
When $\chi^{\prime} = -1$ and simultaneously the imaginary component $\chi^{\prime \prime}$ approaches 0, the sample totally transforms to a superconducting state, and so the external field is completely shielded by the supercurrent around its surface, known as the Meissner effect \citep{Muller}.

\begin{figure}
\centering
\includegraphics[scale=0.29]{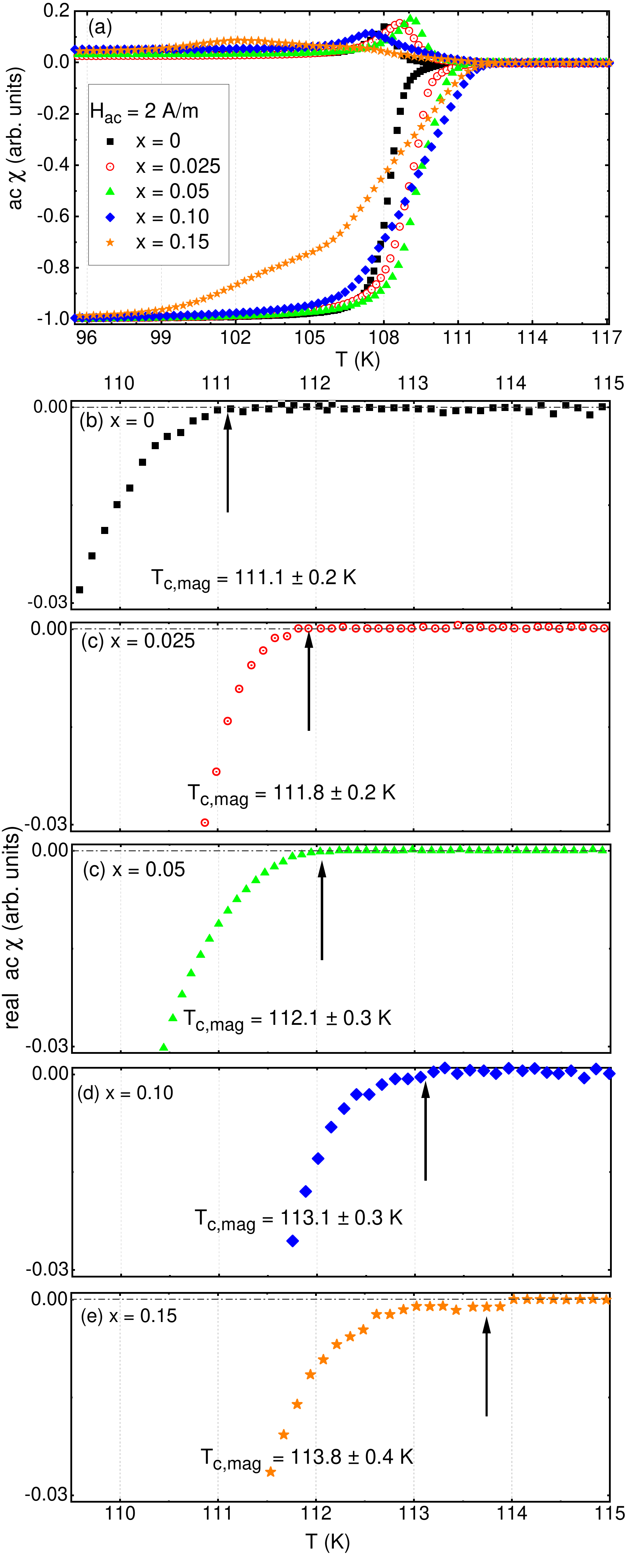}
\caption{ (a) The ac susceptibility for the series of Li-doping in  \chem{Bi_{1.6} Pb_{0.4} Sr_2 Ca_2 (Cu_{1-x} Li_x)_3 O_{10 + \delta}} (with $x = 0.0$, 0.025, 0.05, 0.10, and 0.15) are measured in the temperature range of 95--117 K. 
(b)--(e)  We illustrate in detail the critical transition region of 109.5--113 K; the arrows indicate the critical points, and the dash-dotted lines represent the separation between the paramagnetic and diamagnetic states.  \label{fig:3}}
\end{figure}

At first, we employed the magnetic susceptibility for the $x = 0$ sample at the low magnetic field of $H_\textrm{ac} = 2$ A/m and the frequency of 10$^4$ Hz. 
In Fig.~\ref{fig:2}, we present our notable magnetic susceptibility in the range of 90--115 K along with other references for Bi-2223 compounds prepared by various conditions.
We observe the high $T_\textrm{c, mag}$  and sharp drop with a single step transition in the narrow range of 106.1--111.1 K (green circle in Fig.~\ref{fig:2}), specifically display with $T_\textrm{c,mag} = 111 \pm 0.3$ K in undoped (Bi,Pb)-2223 compound with the $\Delta T$ = 5 K [Table~\ref{table0} and Fig.~\ref{fig:3}(b)].
In particular, no sign of Bi-2212 impurity is emerging in our susceptibility measurement.
That proves the uniformly homogeneous sample of our superconductor.

We juxtapose our magnetic transition with other findings in the Bi-2223 superconductors.
Initially, we observed our critical temperature in (Bi,Pb)-2223 is comparable to the highest value measured in the pristine single crystal with 110.5--111 K, while the transition width is marginally larger than pristine materials (5 K compared with 3 or 4 K in  Refs.~\cite{PhysRevB-O2-hole, Clayton}).
However, even very recent pure single-crystal exemplar shows  $T_\textrm{c,mag} = 110$ K with the $\Delta T = 10$ K, which is longer range transition than our compound  \cite{PhysRevLett-2020-PRL}.
Recently, several newly techniques have been applied to synthesize (Bi,Pb)-2223 compounds such as are nano-coprecipitation with $T_\textrm{c, mag} = 107$ K \cite{nano13152197},  spray-pyrolysis following the postannealing process with $T_\textrm{c, mag} = 109.4$ K\cite{CAO202412212} or our conventional solid-state fabrication with $T_\textrm{c, mag} = 109.5$ \cite{ManP-Ag} (see the list in Table~\ref{table0} and their susceptibilities in Fig.~\ref{fig:2}).
For example, spray-pyrolysis (Bi,Pb)-2223 in Ref.~\cite{CAO202412212} shows reaching of the complete diamagnet below around 95 K even they measure at a low field of 40 A/m  a frequency of 333 Hz (see Exp.~2 plot in Fig.~\ref{fig:2} for the illustration).
So, our result synthesized by the renovated solid-state method demonstrates higher $T_\textrm{c, mag}$ by 1.5--4 K than previous fabricating methods, sharp transition, and approaches the optimal value.   
In order to elucidate this impression,  
 we pioneer the long-time sintering, and three intermediate steps play a pivotal role in 
 fabricating the exceptional superior quality of Bi-2223 (See Table~\ref{table0} and Fig.~\ref{fig:2}.
 An analogous development was recently made by H.\ Cao {\it et al.}~that the doubled pressing and sintering process and promoted a more uniform sample and thus achieved a higher $T_\textrm{c}$ \cite{CAO202412212}.

 Although there has been renowned that lead-substituted (Bi,Pb)-2223 is usually smaller $T_\textrm{c}$  than in  pristine Bi-2223 compound, e.g., decreasing 3 K in the  single crystal demonstration presented in Ref.~\cite{Pb-doped-Bi2223}, 
 our result of (Bi,Pb)-2223 compound draws the opposite attention.
The accomplishment of higher critical temperature in superconductors is greatly enhanced by optimizing the fabrication process even for (Bi,Pb)-2223 compounds.
Especially, 
It takes longer to ensure uniform crystal growth, so the lead-substituted Bi-2223 attains analogously to the pristine compound without a large impact of Pb. 
The large $\Delta T = 5$ K may be due to the ceramic synthesizing method, which creates multisize polycrystalline with different crystal directions (see the microstructure in Fig.~\ref{fig:8}). 
Even in the single crystal growth, attaining 100\% of the high-temperature Bi-2223 phase remains a challenging task, and advanced fabrication techniques are needed to acquire better quality.
Nevertheless, to our knowledge, the maximum $T_\textrm{c}$ of the pure and Pb-substituted Bi-2223 reaches the level of 110--111 K, which might be considered the optimal value of trilayered BSCCO compound.

We investigate the mean of ac susceptibility for the  Li-doped (Bi,Pb)-2223 series at a low magnetic field $H_\textrm{ac} = 2$ A/m to probe the superconducting transitions. 
Fig.~\ref{fig:3}(a) characterizes the smooth curves for most samples for both real (lower) and imaginary (upper) $\chi$ and reach a perfectly diamagnetic state above 105 K, except $x = 0.15 $ case.
The longer-tail $\chi^{\prime}$ transition in heavy Li-doped case may correspond to a higher amount of Bi-2212 matrix or large local distortion in the cuprate plane caused by Li$^+$ cations, 
but it still approaches an absolutely diamagnetic regime at 100 K, which is above the  Bi-2212 maximum transition point of 98 K \cite{Hobou, AC-2023-Bi2212}.
In Fig.~\ref{fig:3}(b)--(e), we depict the pinpoint superconducting transition regime (from 110--115 K) in which  $T_\textrm{c, mag}$ steadily ascends from 111.1 to 113.8 K with respect to the Li-content of $x = 0.0-0.15$.
While $x =0.0$--0.10 samples engage with prompt sharp transition drop from non-superconducting to superconducting phase with a small fluctuation of 0.2-0.3 K, $x = 0.15$ compound exhibits a wider  fluctuation of 1 K around 113.6 K.
The substantially large Li-doping gives rise to multiple phases, including high and low high-$T_\textrm{c}$ (Bi,Pb)-2223 coexisting in the system, which is hardly homogenized.

If we consider the pristine compound with the maximum $T_\textrm{c}$ = 111 K represents  the optimal doped regime, as described in previous studies that increasing hole concentration turns   
our samples into the OD regime in which $T_\textrm{c}$ value is slightly reduced  or behaves like a plateau  \cite{Fujii, PhysRevB-O2-hole, B-Liang-2004, PhysRevLett-2020-PRL}.
However, the critical temperature monotonically augments in the Li-doped (Bi,Pb)-2223 superconductors. 
We suspect that the $T_\textrm{c}$ of 110.5-111 K is not yet an ultimately superlative value of the trilayer Bi-2223.
To illuminate the mechanism driving the $T_\textrm{c}$ increment trend in the Li-doped (Bi,Pb)-2223, 
we propose that the hole concentration is greatly supplemented by replacing a monovalent radius-compatible  Li$^+$ (0.76 \AA) for Cu$^{2+}$ (0.73 \AA) ions.
In pristine (Bi,Pb)-2223, the oxidation state of Cu in the cuprate plane varies from +2 to +3, which is proved by our X-ray photoemission spectrum in (Bi,Pb)-2223 and a recent study \cite{CAO202412212, Man-XPS}.
Introducing monovalent Li$^+$ ions into the cuprate system,  the overall oxidation state of superconducting layer CuO$_2$ tends to reduce close to +2 value.
As shown in the Bi-2223 study in Ref.~\cite{Hole-Bi2223}, the hole concentration is greatly enhanced if the oxidation state of Cu decreases toward +2.
The ubiquitous value of the optimal hole concentration is established with  0.16 \cite{Generic-PRB} in the  homogeneous double-layered Bi-2212 and the YBCO compounds, so 
we consider $p_\textrm{aver} = 0.16$ corresponds to $T_\textrm{c} = 110$ K.
However, to analyze that terminology, we employ  calculating the relation between the hole concentration $p_\textrm{aver}$ (average value of hole-doped content in the CuO$_2$ planes) for several doped Bi-2223 compounds following:
\begin{equation}
    T_\textrm{c} = T^\textrm{max}_\textrm{c} \big [ 1 - 82.6  ( p_\textrm{aver} - p_\textrm{max} )^2 \big ]. 
    \label{eq4}
\end{equation}
$T^\textrm{max}_\textrm{c} = 113.8 $  K denotes the highest up-to-date critical temperature ever observing in the Bi-2223 materials by magnetic measurements. 
And $p_\textrm{max} = 0.18$  is proposed to feature a new optimal hole doping for Bi-2223 superconductors.

\begin{figure}
\centering
\includegraphics[scale=0.34]{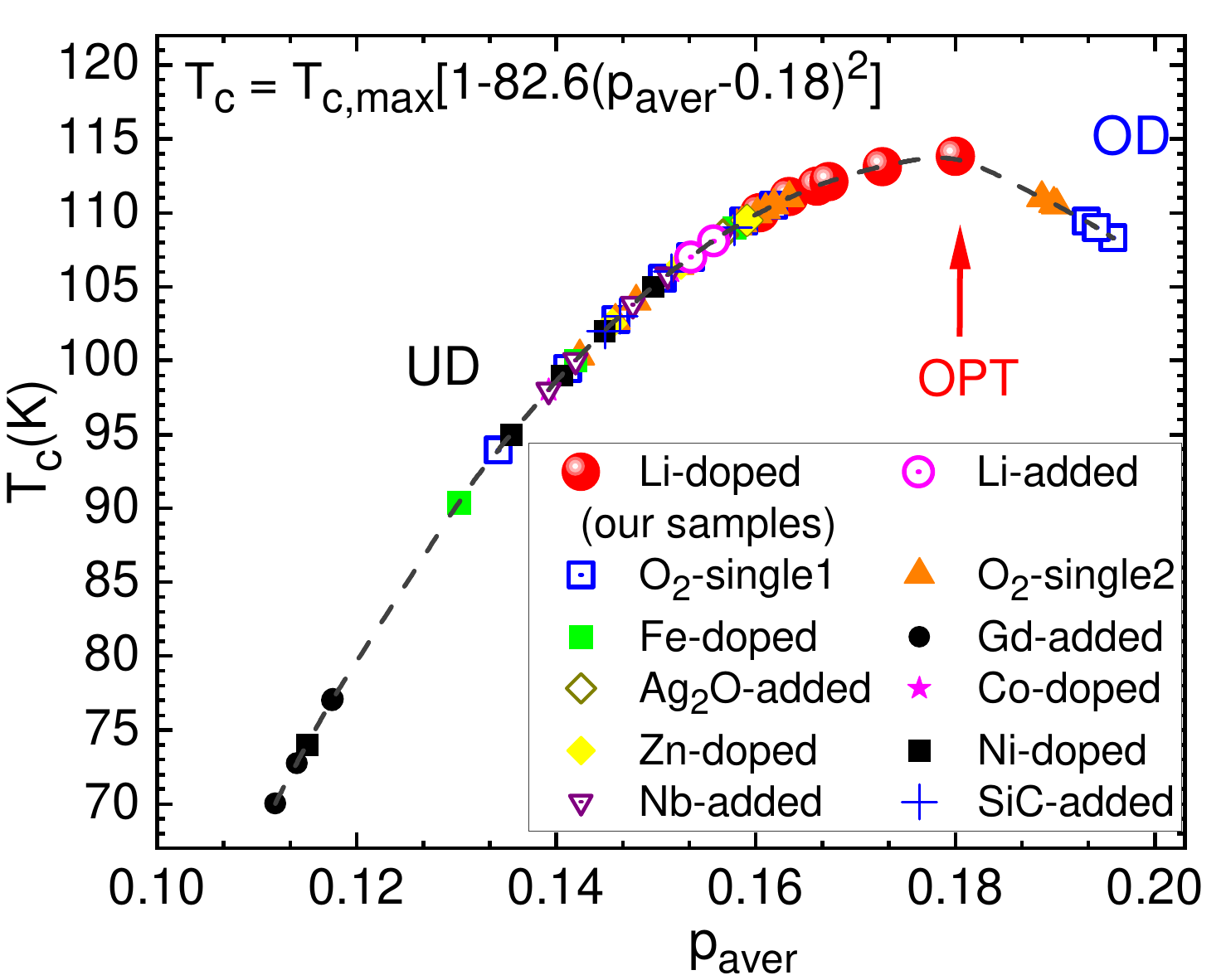}
\caption{ Generic relation of $T_\textrm{c}$ vs hole concentration $p_\textrm{aver}$ in Bi-2223 measured by ac or dc susceptibility and governed by Eq.~(\ref{eq4}). 
We have noted detailed information on the nominal composition and doping concentration.
(i) our Li-doped  samples with composition of  \chem{Bi_{1.6}Pb_{0.4}Sr_2Ca_2(Cu_{1-x}Li_x)_3O_{10 + \delta}} ($x = 0.0-0.15$), 
(ii) Li-added of \chem{Bi_{2}Sr_2Ca_4Cu_6Pb_{0.5}Li_xO_{y}} ($x = 0.0-0.2$) \cite{MatsubaraI},  
(iii) varying O$_2$ content in single crystal growths with formula \chem{Bi_{2}Sr_2Ca_2Cu_3O_{10 + \delta}} through driving the O$_2$ pressure postannealing process  Ref.~\cite{PhysRevB-O2-hole, B-Liang-2004}, 
(iv) Fe-doped compound with formula \chem{Bi_{1.6}Pb_{0.4}Sr_2Ca_2(Cu_{1-x}Fe_x)_3O_{10 + \delta}} ($x = 0 -0.02$) Ref.~\cite{Pop}, 
(v) Gd-added superconductors with  \chem{Bi_{1.7-x}Pb_{0.3}Gd_xSr_2Ca_2Cu_3O_{10 + \delta}} ($x = 0 -0.02$) \cite{Gd-Suscept}, 
(vi) \chem{Ag_2O}-added samples similar to ours \cite{ManP-Ag}, 
(vii) Co-substitution in \chem{Bi_{1.6}Pb_{0.4}Sr_2Ca_2(Cu_{1-x}Co_x)_3O_{10 + \delta}} \cite{Co-doped-Pop} ($x$ = 0.0--0.05), 
(viii) Zn-doped \chem{Bi_{1.6}Pb_{0.4}(Sr_{1.8}Ba_{0.2})Ca_2(Cu_{1-x-y}Zn_xFe_{y})_3O_{10 + \delta}} \cite{Pop-Zn} ($x$ = 0.0--0.02, $y$ = 0.0--0.01), 
(ix) Ni-replaced  \chem{ Bi_{1.6}Pb_{0.4}Sr_2(Ca_{1-x}Ni_{x})_2Cu_3O_{10 + \delta}} \cite{S-Celebi-2004} ($x$ = 0--0.10), 
(x) Nb-doped samples with formula of \chem{Bi_{1.7-x}Pb_{0.3}Nb_xSr_2Ca_2Cu_3O_{10 + \delta}}  \cite{Bilgili-Nb-doped} ($x$ = 0--0.20), 
(xi) SiC-added  \chem{(Bi,Pb)_{2}Sr_2Ca_2Cu_3O_{10 + \delta}} \cite{SiC-added}.
\label{fig:4}}
\end{figure}

As depicted in Fig.~\ref{fig:4}, our Li-doped and Li-added \cite{MatsubaraI} (Bi,Pb)-2223 characterize the raising $T_\textrm{c}$ with respect to the hole concentration.
To obtain a consistent analysis, we take the data measured by magnetic susceptibility for various Bi-2223 compounds.
By controlling the O$_2$ pressure during the postannealing process, the Bi-2223 single crystal can be governed from the UD to OPT and OD regime via the hole concentration from $p_\textrm{aver}$ 0.1--0.2.
On the other hand, previous doping transition metal such as Fe, Ni and Co into the system greatly reduced $T_\textrm{c}$ as well as the hole concentration.
For instance, only substituting 0.02 \% at.~of Fe for Cu, the $T_\textrm{c}$ suppressed from 109 to 90 K that corresponds to the decreasing $p_\textrm{aver}$ from 0.158 to 0.13 \cite{Pop}.
That may be due to Fe usually equips with the oxidation state of +2 or +3 even though the radii of Fe$^{2+}$ and Fe$^{3+}$ are 0.70 and 0.60 \AA, respectively, which are closely matched with Cu$^{2+}$ of 0.73 \AA.
Our results show that the Li-doped (Bi,Pb)-2223 samples are of great quality and are in the optimal regime 111.1-113.8 K range. 
The gap between the OPT (red symbol) and OD regime promises a potential possibility to explore in the future, and the $T_\textrm{c,max}$ can be raised to a higher value or slightly reduced in OD.

From our representation, the $T_\textrm{c}$ versus $ p_\textrm{aver}$ plot shows that the $T_\textrm{c}$ in inhomogeneous triple-layered CuO$_2$ planes analogously follow the trend of Bi-2212, but with higher optimal hole concentration and the OD regime may be lightly dependent on the increasing hole content.
That is well-consistent with the $T_\textrm{c}$ tendency predicted in Refs.~\cite{PhysRevB-O2-hole, PhysRevLett-2020-PRL}.
In the next section, we will combine with resistivity measurements to argue details about the superconducting mechanism.

\subsection{Resistivity measurements, thermodynamic conductivity fluctuation and intrinsic quantities }

\begin{figure}
\centering
\includegraphics[scale=0.285]{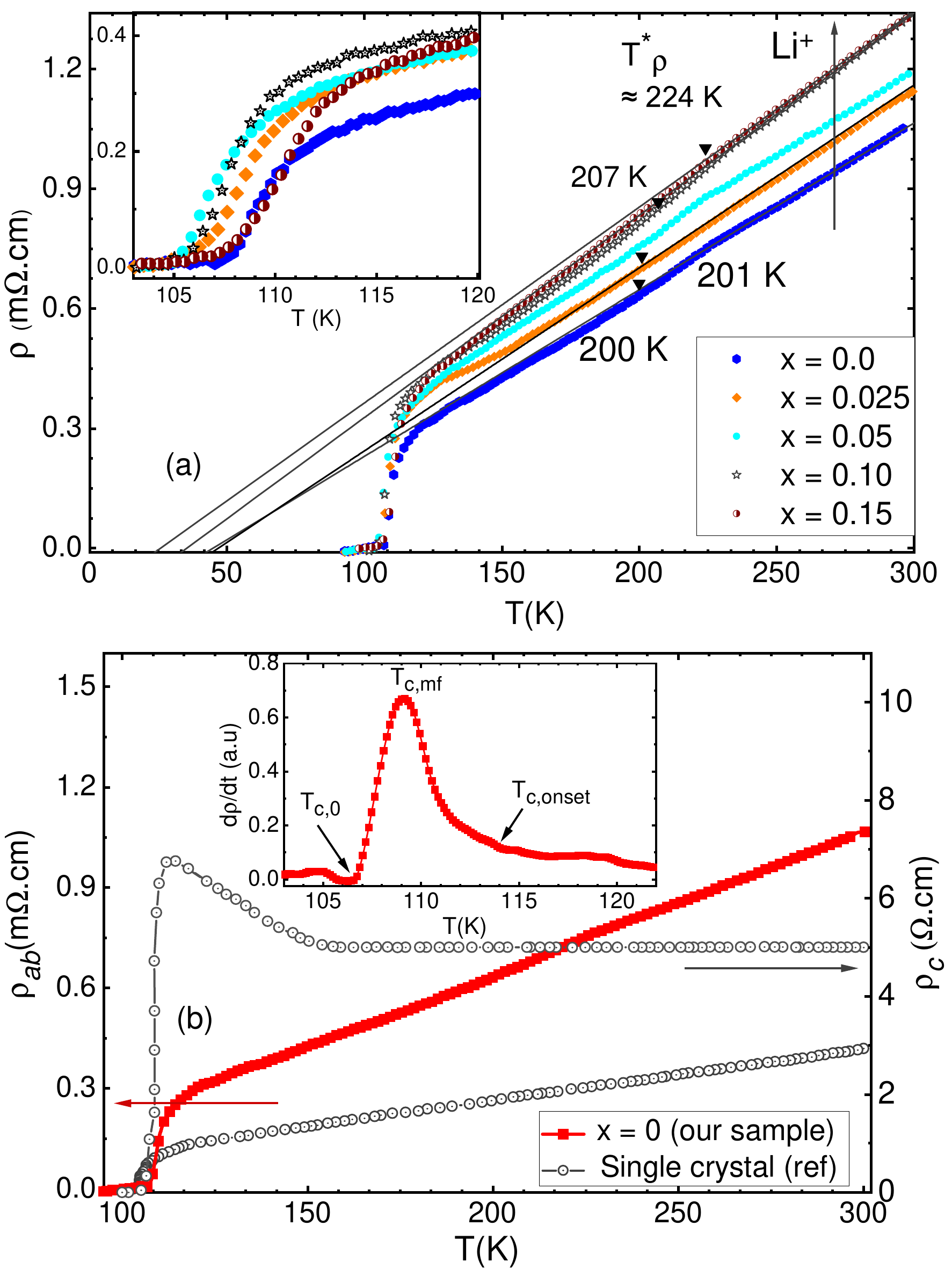}
\caption{ (a) Full resistivity versus temperature $\rho(T)$ of  \chem{ Bi_{1.6}Pb_{0.4}Sr_2Ca_2(Cu_{1-x}Li_x)_3O_{10 + \delta}}   (with $x = 0.0$--0.15) compounds measuring in the temperature range of 85--300 K, and the poor metal-like transition $T^\ast_{\rho}$ show increasing with respect to Li-doped content.
(b) Depict the resistivity of our $x = 0$ sample and the exemplar single crystal obtaining from Ref.~\cite{Fujii} (specimen h). 
The inset illustrates the first derivative of $\rho(T)$ versus temperature. \label{fig:5}}
\end{figure}

\begin{table*}
 \caption{ List the values of the ratio $\rho(300)$/$\rho(120)$ of resistivity at 300 K and 120 K (right before the superconducting transition occurs), $T_\text{c, onset}$  in which the superconducting transformation triggers, $T_\textrm{c, mf}$: mean-field critical temperature, $T_\text{c, 0}$: zero transition temperature, $\rho(120)$/$\rho(105)$ the resistivity ratio at 120 K and complete superconducting phase, zero-temperature coherence length $\xi_c(0)$, effective thickness $d$, Josephson interlayer coupling $\mathcal{J}$ and hole concentrations $p_\textrm{aver}$, $p_\textrm{OP}$ (outer), and $p_\textrm{IP}$ (inner)   of \chem{ Bi_{1.6}Pb_{0.4}Sr_2Ca_2(Cu_{1-x}Li_x)_3O_{10 + \delta}} superconductors ($x$ is the Li-doped content)  \label{table1}}
 \begin{ruledtabular}
  \begin{tabular}{l c c c c c c c c c c c c} 
\small x & $\rho(300)$/$\rho(120)$ & $T_\text{c, onset}$ (K)  & $T_\textrm{c, mf}$ (K) & $T_\text{c, 0}$ (K) & $\rho(120)$/$\rho(105)$ & $\xi_c(0)$ (\AA) & d (\AA) & $\mathcal{J}$ & $p_\textrm{aver}$ & $p_\textrm{OP}$ & $p_\textrm{IP}$ \\ 
 \hline
0 & 3.5 & 113.5   & 109.0 &  107.5 & 44 &  3.44 &   29.90 & 0.053 & 0.163 & 0.188 & 0.114  \\
0.05 & 3.1 & 114.4  & 106.2 &  104 & 78 & 4.07 &  36.76 &  0.049 & 0.167 & 0.192 & 0.118\\
 0.10 & 3.2 & 116.5 & 110.5 & 105.0   & 100 &  4.22 & 37.61 &  0.050 & 0.173 & 0.197 & 0.123\\
0.15 & 3.2 & 117.3 & 110.8 & 105.0  & 100 &  4.93 & 45.03 &   0.048 & 0.180 & 0.205 & 0.131 \\
\end{tabular}
 \end{ruledtabular}
\end{table*}

 The temperature dependence of resistivity $\rho(T)$ for the Li-substituted materials in the range of 85--300 K is exhibited in Fig.~\ref{fig:5}(a). 
At the normal state $T > T^{\ast}_{\rho} $, the resistivity $\rho$  linearly behaves with respect to temperature following a relation $\rho_\textrm{n}(T) = \rho_0 + aT$ in which the residual resistivity $\rho_0$ and slope $a$ are determined from the plot.
These parameters will be utilized to calculate the excess conductivity fluctuation. 
The $\rho(T)$  at the normal state regime increases monotonically from $x = 0.0$ to $x = 0.15$, [an arrow in Fig.~\ref{fig:5}(a) indicates raising \chem{Li^+} content]. 
Our Li-doped (Bi,Pb)-2223 materials demonstrate a  negative residual resistivity, which is well-matched with the prior investigation in the Bi-2223 single crystal \cite{Fujii}.

Around  the OPT region, increasing the hole concentration reduces the  pseudogap  transition versus $T^{\ast}$  as observed in several reports \cite{Keimer,Fujii}.
However, in the Li-doped (Bi,Pb)-2223, 
 $T^{\ast}$ proportionally increase versus the hole concentration.
 So, the linear region may not correspond to the pseudogap, 
 rather it belongs to the poor metal-like behaviors with the resistivity at room temperature around 1 m$\Omega$.cm which is a thousand times larger than the typical resistivity of normal metals ($\rho_\textrm{Cu} = 1.68\times 10^{-3}$ m$\Omega$.cm).
Overall slopes $d\rho /dT$ monotonically
raises following Li-doping, 
whereas the ratio  $\rho(300)/\rho(120)$ almost remains 3.2-3.5 (Table~\ref{table1}).
This immediately suggests that  all samples belong to the same class of materials above the critical transition. 
At the temperature $T^{\ast}$, the Cooper pairs may already start to form, while phase fluctuations prohibit superconducting order until lower temperatures
 \cite{Keimer, NaturePhys-2023-Bi2223}. 
The analogous observation was proposed in Li-doped \chem{YBa_2Cu_{1-x}O_{7 + \delta}Li_x} (YBCO) compounds \cite{Sauv,Bobroff}. 
The substitution of \chem{Li^+} impurities   for \chem{Cu^{2+}} cations on the cuprate plane behaves like scattering centers that impede current flow in the non-superconducting phase.

We discuss the Li-doping effect on the critical temperatures of (Bi,Pb)-2223 phase.
 Fig.~\ref{fig:5}(a) and its inset depict our ceramic compounds as fully superconductive above 105 K, which is dominated by high-$T_\textrm{c}$ (Bi,Pb)-2223 phase.
Since the $\rho(T)$ is difficult to justify correctly the transition point, 
the first derivative $d\rho/dT$ is evaluated to concretely provide the phase transition points [Fig.~\ref{fig:5}(b)].
Increasing Li-doping content raises $T_\text{c, onset}$ about 4 K from 113.5 K at $x = 0$  to 117.3 K at $ x = 0.15$, [see Table.~\ref{table1} and inset of Fig.~\ref{fig:5}(b)].
The signature of forming the Cooper pair in the superconducting state is clearly enhanced overall with hole concentration, which is consistent with the previous investigation by the magnetic susceptibility.
On the contrary, the prior resistivity measurements of Li-doped samples \cite{Li-doped-Bi2223,Li-Bi2223} indicated no change of $T_\textrm{c}$ or having variation in the range of 107-110 K. 
That may be due to their samples not being optimized in the fabrication of the undoped samples before doping. 
In other examples, doping strong magnetic rare-earth such as Nd and  Gd into Bi-2223 host strongly deteriorates $T_\text{c}$ and transport properties \cite{ Terzioglu, DOGRUER2022126350}. 

\begin{figure}
\centering
\includegraphics[scale=0.173]{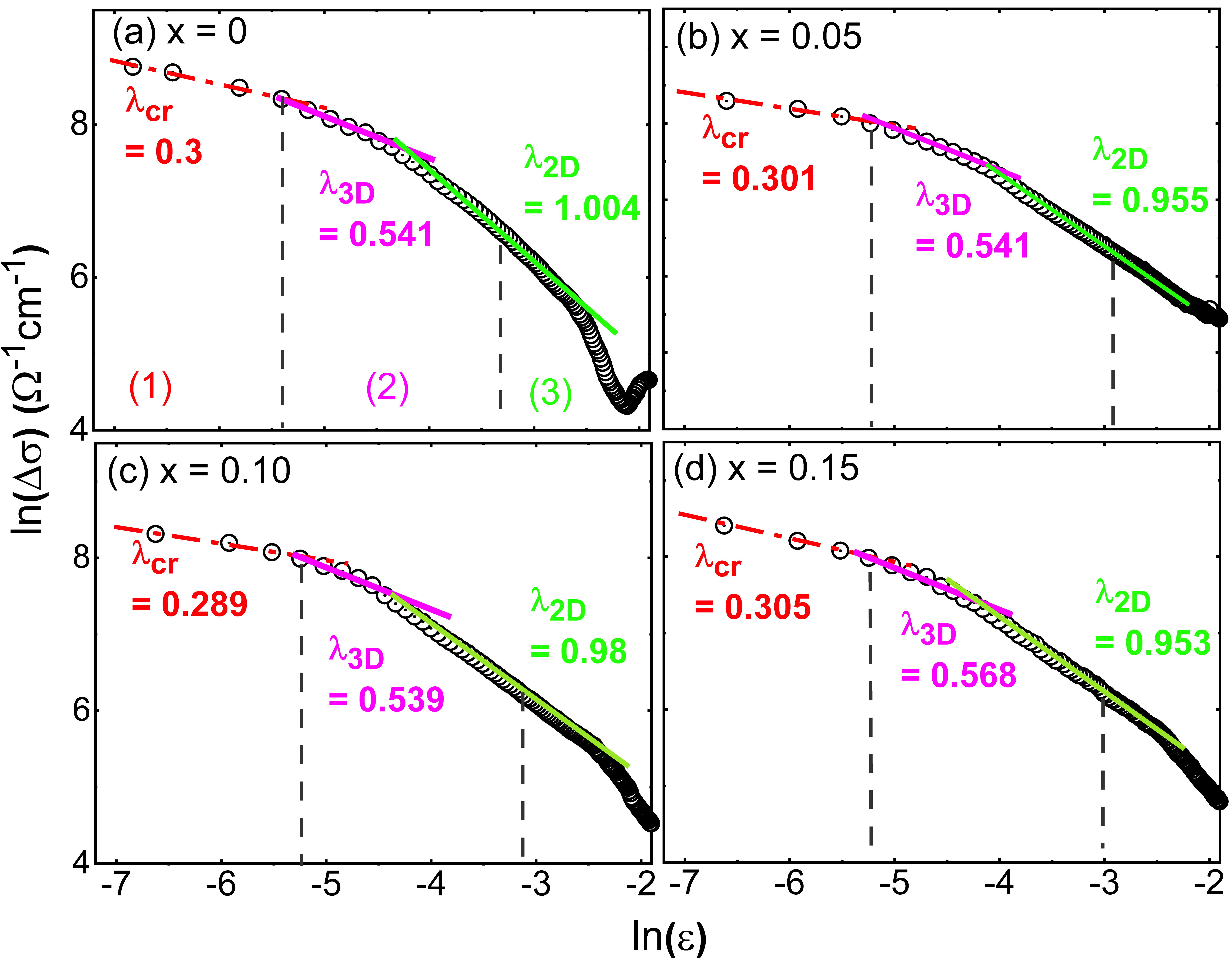}
\caption{ Double logarithmic plot of the excess conductivity fluctuation $\Delta \sigma$ as a function of the reduced temperature $\epsilon = (T - T_\textrm{c,mf})/T_\textrm{c,mf}$ for \chem{ Bi_{1.6}Pb_{0.4}Sr_2Ca_2(Cu_{1-x}Li_x)_3O_{10 + \delta}} with (a) $x = 0$, (b) $x = 0.05$, (c) $x = 0.10$ and (d) $x = 0.15$, respectively. 
The labels of (1), (2) and (3) are referred to the critical
region (CR), the mean-field region (MFR)
and the short wave fluctuation (SWF), respectively, which  are separated by the dashed lines. \label{fig:6}}
\end{figure}

From the normal resistivity in the above part, we calculate the  
logarithm of the excess conductivity $\ln (\Delta \sigma)$ as a
function of the reduced temperature $\epsilon$ for our representative samples (x = 0.0--0.15) 
using Eqs.~(\ref{eq0})--(\ref{eq01}), and we display in Fig.~\ref{fig:6}(a)--(d). 
The overall shape of $\ln (\Delta \sigma)$--$\ln (\epsilon)$ features analogously in all Li-doped samples. 
The critical exponent  parameters $\lambda$ are fitted following the AL model and delve into 
the critical
region (CR), the mean-field region (MFR)
and the short wave fluctuation (SWF) as the temperature turns away from the critical point $T_\textrm{c, mf}$. 
The dynamical CR of the samples with the exponents around $\lambda = 0.3$ (Table~\ref{table1}) in 
which the Landau-Ginzburg theory is broken down \cite{PhysRevB-Lobb}.
Both values of 0.289--0.305 are in great agreement with the
theoretical conjecture of the dynamic scaling effect \cite{PhysRevB-Lobb}, and that  proves 
the consistency and accuracy of the LD theory for our analysis in the Li-doped trilayered (Bi,Pb)-2223. 

Particularly, the  critical exponent in the MFR 
is around $\lambda \sim 1$ which represents the 2D conductivity fluctuations and that  mainly occurs forming of the 
Cooper pairs within the CuO$_2$ layers \cite{Law-Do, Super-Fluc-Bi2223}. 
The MFR region demonstrates a pronounced effect resulting from Li-doping that a narrow range in the updoped sample [Fig.~\ref{fig:6}(a)] drastically drops with $\lambda$ larger than 1 to a wider range in the doped samples [Fig.~\ref{fig:6}(b)--(d)] with all $\lambda$ magnitude smaller than 1.
It is implied that the Li-doped samples signify a slight appearance of 3D fluctuation between the CuO$_2$ planes even above the critical transition.
As temperature 
closely approaches  to $T_\textrm{c,mf}$, $\lambda$ reduces to $\lambda = 0.5$ (3D fluctuation).
At this temperature range, the Cooper pairs strongly tunnel through insulating CaO layers by the Josephson effect to
reach the conducting CuO$_2$ layer along $c$-axis \cite{Law-Do}.
Our findings are extremely harmonious with the observation of the Josephson tunneling effect of the Cooper pairs between IP and OP in the OPT and OD regions \cite{PhysRevB-O2-hole, PhysRevLett-2020-PRL}.

Fig.~\ref{fig:5}(b) illustrates our undoped (Bi,Pb)-2223 with an exemplar resistivity measurement on single crystal along the $ab$-plane and $c$-axis  \cite{Fujii}.
Above the critical transition, our (Bi,Pb)-2223 superconductors favor conducting along the $ab$--plane with the magnitude about a half of the single crystal.
Along the $c$-axis, the resistivity is a thousand times larger than on the $ab$-plane. 
In other words, the $c$-axis
electronic conductivity is very weak and within the incoherent limit at the normal phase.
However, around the critical point, both types of transports can simultaneously conduct.
However, the coherent $c$-axis transport suddenly occurs below $T_\textrm{c}$ 
due to tunneling of the Josephson currents across the insulating
space layers \cite{B-Liang-2004}.
That proves the precise emergence of the 3D-2D transition, which we identify in the thermodynamic conductivity analysis.

 The effective layer thickness $d$ and zero-temperature coherence length $\xi_c(0)$  are computed from 2D and 3D critical components by Eq.~(\ref{eq1}), respectively, from the intercept quantities of the linear fitting. 
 Table~\ref{table1} presents 
the $\xi_c(0)$ and $d$ steadily increased 3.44--4.93 \AA~and 29.90--45.03 \AA, respectively. 
 $\xi_c(0)$ is comparable to the distance between the IP and OP cuprate planes of 3.3 \AA~in Fig.~\ref{fig:1}, 
 whereas the the effective layer thickness $d$ is around the length of Bi-2223 $c$-axis of 37.13 \AA~\cite{Rainer}.
Also, the upper critical magnetic field is inferred from the coherent length with the magnitude of larger than 1000 T \cite{Rainer}.
$T_\textrm{c}$ is believed to be related to the coherence length \cite{OH2019348, exc-conduct-Bi-2223}, 
and thus explain the enhancement of $T_\textrm{c}$ in the Li-doped samples.
However, our correlation length is about half of single crystal Bi-2223 with 10 \AA~in 
\cite{PhysRevB-O2-hole, Pb-doped-Bi2223}  inferred from a magnetic measurement.
This may be due to better quality along the $c$-axis in single crystal growth in compared with our ceramic compounds or related to the magnetic Landau-Ginzburg  coherence length rather than Josephson coherence length.
Our result suggests a purposeful possibility to fabricate a Bi-2223  monolayer like Bi-2212 for further investigation \cite{mono-layer}.

The interlayer coupling $\mathcal{J}$  calculated from Eq.~(\ref{J-coupling}), retain almost unchanged with respective to Li-doping  (Table~\ref{table1}).
The Josephson tunneling effect $\mathcal{J}$ between CuO$_2$ planes demonstrates much weaker than the effective superexchange coupling between Cu-O-Cu on cuprate planes about a third or a six of OP and IP, respectively   \cite{Science-2023-Bi2223}.
Maintaining the Josephson tunneling of
the Cooper pairs between the inequivalent CuO$_2$ planes within
a unit cell has also been proposed to be crucial for
enhancing $T_\textrm{c}$. 
Therefore, in Bi-2223 compounds, the electronic transport is strongly preference in the $ab$-plane because of a hybridization coupling between Cu-$d_{x^2-y^2}$ and O-$2p_x/p_y$ \cite{Keimer, NaturePhys-2023-Bi2223} and crystal stacking growth (see Fig.~\ref{fig:8}).

Provided that the Josephson coupling parameter remains nearly constant in the Li-doped (Bi,Pb)-2223, the hole concentrations in IP and OP retain unchanged \cite{hole-content1-PRB, hole-content2, hole-content3}.
 OP preserves the hole content in OD while IP keeps in the the UD regime because of the protection of two insulating layers, CaO and the cleanest sheet among the cuprate planes.
The hole difference quantity $\Delta p = 0.076$ between IP and OP was detected by the Raman scattering and nuclear magnetic resonance at $T_\textrm{c}$ = 109 K \cite{hole-content1-PRB, hole-content2}.
However, since our $T_\textrm{c}$ = 111 K, the value reduces to 0.074 and maintains  through all Li-doped samples.
Accompanied with our $p_\textrm{aver}$ listed in Table~\ref{table1}, the  hole concentrations in IP and OP are calculated.
As $T_\textrm{c}$ increases, the hole content in both the cuprate layers raises as well.
The existence of inequivalent
IP and OP hole concentrations, and other intrinsic properties such as superconducting gap, and superexchange coupling in the CuO$_2$ planes result in the anomalous properties in Bi-2223 as well as promote the highest $T_\textrm{c}$ among BSCCO family with a number of CuO$_2$ $n= 3$.
Our results are consistent with the theoretical proposal \cite{Emery} and several experimental artifacts, which have been recently studied  in the Bi-2223 superconductors  \cite{Statt, Ideta, NaturePhys-2023-Bi2223, Science-2023-Bi2223, PhysRevLett-2020-PRL}.

\subsection{The field dependence of susceptibility and intergrain current density}

\begin{figure}
\centering
    \includegraphics[width=.47\textwidth]{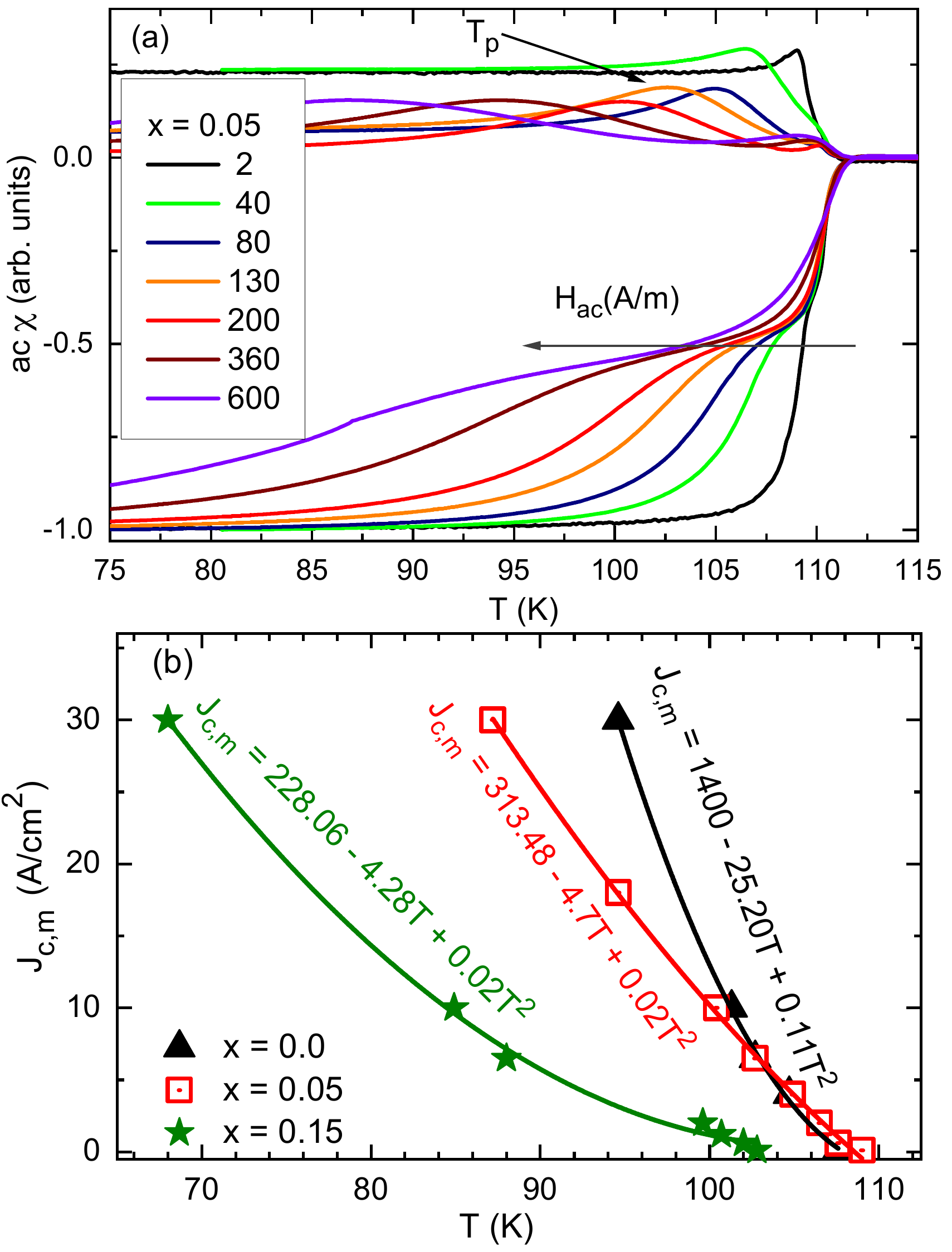} 
 \caption{ (a) Temperature dependence of the complex ac susceptibility ($\chi = \chi^{\prime} + i \chi^{\prime \prime}$) at various magnetic fields and a frequency $f = 1$ kHz for  $x = 0.05$ sample. (b) $J_{\text{c,m}}$ versus $T_{\text{p}}$ for $x = 0$ (black),  $x = 0.05$ (red), and  $x = 0.15$ (green) samples.} \label{fig:7}
\end{figure}

\begin{figure*}
\centering
\includegraphics[scale=0.55]{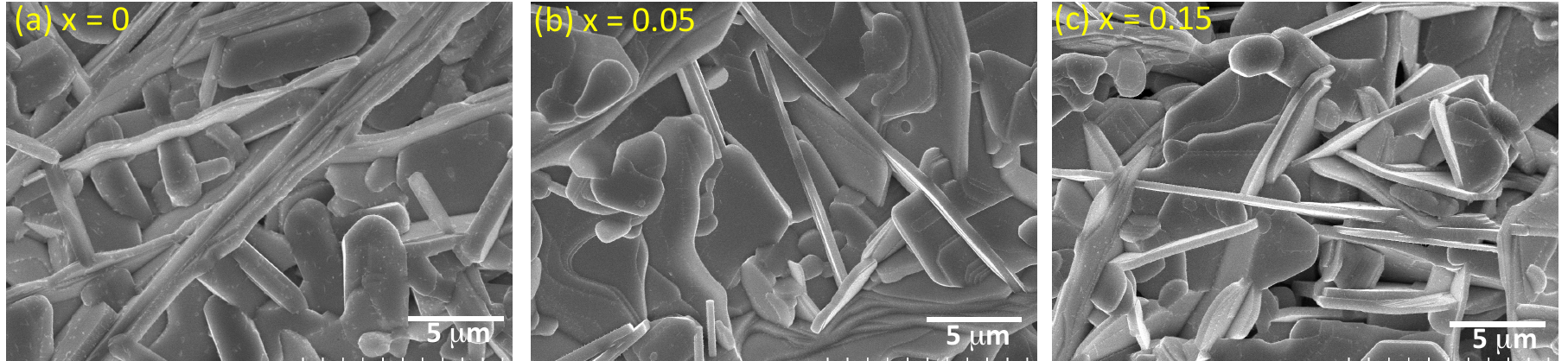}
\caption{ SEM images of (a) $x = 0.0$ (long-range particle range from 2 to 10 $\mu$m, average size of 3.8 $\mu$m) (b) $x = 0.05$ (c) $x = 0.15$ of \chem{ Bi_{1.6}Pb_{0.4}Sr_2Ca_2(Cu_{1-x}Li_x)_3O_{10 + \delta}} ceramic compounds \label{fig:8}}
\end{figure*}

The ac susceptibility is a highly adaptable method which characterize the critical temperature  and vortex dynamics of superconductors.
%
Figure \ref{fig:7}(a) demonstrates the physical effect of ac magnetic field $H_{\text{ac}} = 2-600$ A/m on $x = 0.05$ sample in the temperature range of 75--115 K, at the frequency $f = 1$ kHz. 
Above the high magnitude of 80 A/m,  the full $\chi(T)$ curves clearly show  a two-step transition in the real part $\chi^{\prime}$ with a long tail, and the upper imaginary part $\chi^{\prime \prime}$ unveils the ac dissipation peak. 
The peak of $\chi^{\prime \prime} (T)$ curve at temperature $T_\textrm{p}$ is originally generated from the hysteresis loss of the magnetic cycle of intergrain [Fig.~\ref{fig:7}(a)]. 
The $\chi^{\prime \prime}(T)$ plots display two peaks of intragrain and intergrain  \cite{Muller}. 
The intragrain peaks remain stable due to a strong pinning force giving rise to the  Abrikosov vortices in which the mixed state with partly magnetic flux penetrates through a non-ideal type-II superconductor. 
The intergrain peak $T_\text{p}$ shows the weak-link between superconducting grains representing the Josephson vortices. 
When the magnetic amplitude increases, the broader intergrain peaks indicate higher ac loss energy \cite{Muller}. 
They demonstrate that the weak-link coupling between superconducting grains of multiphase sample makes $J_\text{c,m}$ lower and strongly sensitive to magnetic field \cite{Larbalestier}.

The intergrain matrix current density $J_\textrm{c,m}$ is computed by utilizing the peaks of imaginary $\chi^{\prime \prime} (T)$ by the Bean critical model \cite{Bean}. 
According to the model of the bar-shaped sample, the $J_{\text{c,m}}$ at the peak temperature $T_{\text{p}}$ is calculated by formula: $J_{\text{c,m}} = H_{\text{ac}}/(W\times D)^{\frac{1}{2}}$, where the cross section of the rectangular bar-shaped sample denotes $2W\times 2D$ (provided in Sec.~\ref{Exp}). 
$J_\textrm{c,m}$ value strongly depends on the external ac magnetic field, temperature, and slightly changes with frequency [see Figs.~\ref{fig:3}(a) and \ref{fig:7}(a) for a comparison].      
 Figure \ref{fig:7}(b) manifests the mean of  the matrix critical current density $J_{\textrm{c,m}}$ as a function of temperature. 
 When the $\chi^{\prime \prime}-T$ broadens at $T_{\text{p}}$, the pinning force is weaker, $J_{\text{c,m}}$ value is smaller. 
 Using a power-law curve fit, the $J_{\text{c,m}}$ functions behave in the quadratic forms, 
 and the magnitude of $J_{\text{c,m}}$ are decreased of 163, 70 and 30 A/cm$^2$ (for $x = 0.0$, 0.05 and 0.15, respectively) at 77 K (the boiling point of liquid nitrogen). 
 Analogous to the effect of Li$^+$ in the non-superconducting state, the existence of external impurities Li$^+$ cations in the matrix weak-link phase reduces the transport properties even in the superconducting state.

Fig.~\ref{fig:8} shows the SEM images of undoped and doped (Bi,Pb)-2223 compounds.
The ceramic compounds demonstrate the granular crystal structure with random grain orientations. 
Within the scale of 5 $\mu$m, their surface morphology indicates the strong dark-gray color of micro-size crystal plates (5 - 10 $\mu$m) and disks (3 - 5 $\mu$m). 
The thin plates show the preferential growth of \chem{CuO_2} plane along the $c$-axis packed to form the crystal Bi-2212 matrix and Bi-2223 grain \cite{Cai}. 
Increasing Li-dopant makes samples with a higher number of disks denser as well, and the average grain size increases around 4 $\mu$m (undoped sample) to 7-8 $\mu$m (Li-doped compounds).
The bigger grain size belongs to lower temperatures forming the microparticles in the Li-doped samples \cite{Matsubara}.
That explains the effect on the transport properties in Li-doped compounds.
 
\section{Conclusions \label{conclude}}

We present our renovated solid-state reaction method, which integrates the long-time sintering and three intermediate calcining and pressing stages to successfully fabricate the supreme quality of the Li-doped  (Bi,Pb)-2223 ceramic compounds. 
The ac magnetic susceptibility demonstrates the highest up-to-date $T_\textrm{c,mag}$ ever obtained in Bi-2223 superconductors from 111.1--113.8 K with the Li-doping content of 0-0.15 \% at.
These values overcome the standard $T_\textrm{c} = 110$ K of the Bi-2223 compound.
First, the high $T_\textrm{c}$ is driven by the high-quality homogeneity of samples as well as the enhancement of hole concentration in the CuO$_2$ planes.
The hole concentration is to achieve the new higher optimal value of 0.18 compared with the ubiquitous value in Bi-2212 and other cuprate superconductors of 0.16.
Second, the strong correlation effect between underdoped IP and overdoped OP  \chem{CuO_2} planes enhances the phase stiffness and $T_\textrm{c}$.

The radius-compatible Li$^+$ (0.76 \AA) substituting for Cu$^{2+}$ (0.73 \AA) affects the transport properties in both the non-superconducting and superconducting phases.
The Li-doped samples exhibit the larger resistivity in the normal states and the lower critical current density $J^{\text{c,m}}$ in the superconducting state. 
The \chem{Li^+} ions at \chem{CuO_2} plane are like scattering impurities to impede the current flow and replace Cu-O-Cu with Li-O-Cu interaction.

By rigorously addressing the excess thermodynamic conductivity data near the transition temperature, we observe the zero temperature interlayer coherence length $\xi_c(0)$ increases with respective to the hole-doped content, whereas the Josephson coupling between the cuprate IP and OP  remains unchanged.
The coherence length in the ceramic compounds is comparable to the distance size between IP and OP, and its magnitude is half of obtaining value in the single crystal.
Especially, the Li-doped sample displays larger 3D Cooper pairing fluctuation than in the undoped (Bi,Pb)-2223 compound due to longer coherence length.
Our Josephson interlayer coupling between IP and OP is much smaller than the superexchange Cu-O-Cu coupling inside the CuO$_2$ layer, and thus, it is interpreted as a weak correlation between the cuprate planes.
And so, the transport in Li-doped (Bi,Pb)-2223 is dominated in the $ab$-plane rather than along $c$-axis.
That is the predominant objective to explain the anomalous properties and the highest $T_\textrm{c}$ among the BSCCO family.

Our findings shed light on future investigations on the heavier Li-doping in the system or improving fabrication  methods such as a wire form or a single crystal for the  current Li-doped samples.
That may efficiently and effectively apply for practical applications including bullet train, fusion reaction, magnetic resonance image and other superconducting magnets utilizing the high critical temperature and giant upper critical magnetic field.

\section{Acknowledgment}
Man NK acknowledges Vietnam's National Foundation for Science and Technology Development (NAFOSTED) under grant No: 103.2-2022.
 Huu Do thanks for the financial support from the University of Illinois Chicago. This work was financially supported by the HUST Science and Technology Project Code: T2018-PC-221.

\section{References}

\bibliography{mybibfile}


\nocite{apsrev42Control}
\end{document}